\documentclass[12pt]{article}
\usepackage{graphicx}

\newcommand\pubnumber{SNSN-323-63}
\newcommand\pubdate{\today}

\def\lnf{INFN Laboratori Nazionali di Frascati, I-00044 Frascati, ITALY}
\def\support{\footnote{for the FlaviaNet Working Group on Kaon Decays}}

\def\Title#1{\begin{center} {\Large #1 } \end{center}}
\def\Author#1{\begin{center}{ \sc #1} \end{center}}
\def\Address#1{\begin{center}{ \it #1} \end{center}}

\newcommand\pubblock{\rightline{\begin{tabular}{l} \pubnumber\\
         \pubdate  \end{tabular}}}
\newenvironment{Abstract}{\begin{quotation}  }{\end{quotation}}
\newenvironment{Presented}{\begin{quotation} \begin{center} 
             PRESENTED AT\end{center}\bigskip 
      \begin{center}\begin{large}}{\end{large}\end{center} \end{quotation}}

\def\beq{\begin{equation}}
\def\eeq#1{\label{#1}\end{equation}}
\def\eeqn{\end{equation}}

\def\beqa{\begin{eqnarray}}
\def\eeqa#1{\label{#1}\end{eqnarray}}
\def\eeqan{\end{eqnarray}}

\let\bar=\overbar

\def\Dslash{\not{\hbox{\kern-4pt $D$}}}
\def\dslash{\not{\hbox{\kern-2pt $\del$}}}

\def\msb{{\bar{\ssstyle M \kern -1pt S}}}

\usepackage{amsmath}
\usepackage{amssymb}

\newcommand{\Fig}[1]{Fig.~\ref{#1}}
\newcommand{\Tab}[1]{Table~\ref{#1}}
\newcommand{\BR}[1]{\ensuremath{{\rm BR}(#1)}}
\newcommand{\dEM}[1]{\ensuremath{\delta_{\rm EM}^{#1}}}
\newcommand{\dSU}[1]{\ensuremath{\delta_{\rm SU(2)}^{#1}}}
\newcommand{\order}[1]{\ensuremath{\mathcal{O}(#1)}}

\newcommand{\SU}[1]{\ensuremath{\rm SU(#1)}}
\newcommand{\fp}{\ensuremath{f_+(0)}}
\newcommand{\FK}{\ensuremath{f_K}}
\newcommand{\FKpi}{\ensuremath{f_K/f_\pi}}
\newcommand{\Fpi}{\ensuremath{f_\pi}}
\newcommand{\Vub}{\ensuremath{|V_{ub}|}}
\newcommand{\Vud}{\ensuremath{|V_{ud}|}}
\newcommand{\Vus}{\ensuremath{|V_{us}|}}
\newcommand{\Vusd}{\ensuremath{|V_{us}/V_{ud}|}}
\newcommand{\Vusf}{\ensuremath{|V_{us}|f_+(0)}}

\begin{document}
\begin{titlepage}
\pubblock

\vfill
\Title{$V_{us}$ and precise Standard Model tests}
\vfill
\Author{Barbara Sciascia\support}
\Address{\lnf}
\vfill

\begin{Abstract}
The recent significant progress on both the experimental and theoretical sides
on the study of leptonic and semileptonic kaon decays
allows to precisely test the Standard Model. Here we present results for
the determination of \Vus\ from experimental data, the comparison between 
the values of \Vus\ obtained from data on $K\to\pi\ell\nu$ ($K_{\ell3}$) 
and $K\to\mu\nu$ ($K_{\mu2}$) decays, and tests of lepton universality in $K_{\ell3}$ decays.
\end{Abstract}
\vfill
\begin{Presented}
CKM2010, the 6th International Workshop on the CKM Unitarity Triangle\\
University of Warwick, UK, 6--10 September 2010
\end{Presented}
\vfill
\end{titlepage}
\def\thefootnote{\fnsymbol{footnote}}
\setcounter{footnote}{0}

A detailed analysis of precise tests of the Standard Model 
with leptonic and semileptonic kaon decays has already 
been presented in \cite{Antonelli:2008jg}. 
However, the recent significant progress on both the experimental and theoretical 
sides has motivated us to perform an updated analysis with three major areas
of emphasis: the determination of \Vus\ from experimental data, 
with and without imposing CKM unitarity; the comparison between 
the values of \Vus\ obtained from data on $K\to\pi\ell\nu$ ($K_{\ell3}$) 
and $K\to\mu\nu$ ($K_{\mu2}$) decays and the corresponding 
constraints on deviations from the $V-A$ structure of the charged current;
tests of lepton universality in $K_{\ell3}$ decays.
The complete work can be found in Ref \cite{Antonelli:2010yf}; here we report only on the
main physics results.

Within the Standard Model (SM), leptonic and semileptonic kaon decays
can be used to obtain the most accurate determination of the magnitude of
the element $V_{us}$ of the Cabibbo-Kobayashi-Maskawa (CKM) 
matrix. 
A detailed analysis of these processes potentially also provides
stringent constraints on new physics scenarios:
while within the SM, all $d^i \to  u^j \ell \nu$ transitions 
are ruled by the same CKM coupling $V_{ji}$ (satisfying
the unitarity condition $\sum_k |V_{jk}|^2 =1$), and 
$G_F$ is the same coupling that governs muon decay, 
this is not necessarily true beyond the SM. 
New bounds on violations of CKM unitarity and lepton universality 
and deviations from the $V-A$ structure translate into significant 
constraints on various new-physics scenarios. Alternately, such tests may
eventually turn up evidence of new physics. 
In the case of leptonic and semileptonic kaon decays, these tests 
are particularly significant given the large amount of
data recently collected by several experiments,
the substantial progress recently made in evaluating 
the corresponding hadronic matrix elements from lattice QCD, and
the precise analytic calculations of radiative 
corrections and isospin-breaking effects recently performed
within chiral perturbation theory, the low-energy effective 
theory of QCD.

An illustration of the importance
of semileptonic kaon decays in testing the SM 
is provided by the unitarity relation 
$\Vud^2 + \Vus^2 + \Vub^2 = 1 + \Delta_{\rm CKM}$.
Here the $V_{ji}$ are the CKM elements as determined from 
the various $d^i \to  u^j$ processes, where the value of $G_F$ is determined 
from the muon life time: 
$G_\mu = 1.166371 (6) \times 10^{-5} {\rm GeV}^{-2}$. 
$\Delta_{\rm CKM}$ parameterizes possible 
deviations from the SM induced by dimension-six operators, 
contributing either to muon decay or to $d^i \to  u^j$ 
transitions \cite{Cirigliano:2009wk}. 
The present accuracy on \Vus\ allows us  
to set bounds on $\Delta_{\rm CKM}$ around $0.1\%$,
which translate into bounds on the effective scale of new physics
on the order of 10~TeV.

For each of the five decay modes for which rate measurements exist, we use 
$$
\Gamma_{K_{\ell 3}} = 
\frac{G_F^2m_K^5}{192\pi^3}\, C_K^2 S_{\rm EW} 
\left(\Vus f_+^{K^0 \pi^-}(0) \right)^2 I_{K\ell}
\times \left(1 + \dEM{K\ell} + \dSU{K\pi} \right)^2
$$
to evaluate \Vusf\ from the decay rate data, 
the phase space integrals from dispersive fits, 
the long-distance radiative corrections, 
and the \SU{2}-breaking corrections. 
 We keep track of the 
correlations between the uncertainties on the values of \Vusf\ from different 
modes arising from the use of common corrections and from correlations in 
the input data set. 
The resulting values of \Vusf\ are listed in \Tab{tab:Vusf0}.
\begin{table*}
\centering
\begin{tabular}{lcccccc|rrrr}
\hline\hline
Mode                & \Vusf\       & \% err & BR   & $\tau$ & $\Delta$ & Int & \multicolumn{4}{c}{Correlation matrix (\%)} \\
\hline
$K_L \to \pi e \nu$     & 0.2163(6)    & 0.26   & 0.09 & 0.20  & 0.11  & 0.06 & $+55$ & $+10$ & $+3$ & $0$   \\
$K_L \to \pi \mu \nu$   & 0.2166(6)    & 0.29   & 0.15 & 0.18  & 0.11  & 0.08 &       & $+6$  & $0$  & $+4$  \\
$K_S \to \pi e \nu$     & 0.2155(13)   & 0.61   & 0.60 & 0.03  & 0.11  & 0.06 &       &       & $+1$ & $0$   \\
$K^\pm \to \pi e \nu$   & 0.2160(11)   & 0.52   & 0.31 & 0.09  & 0.40  & 0.06 &       &       &      & $+73$ \\
$K^\pm \to \pi \mu \nu$ & 0.2158(14)   & 0.63   & 0.47 & 0.08  & 0.39  & 0.08 &       &       &      &       \\
Average                & 0.2163(5)  &    &  &   &   & \\
\hline\hline 
\end{tabular}
\caption{Values of \Vusf\ as determined from each kaon decay mode, 
with approximate contributions to relative uncertainty (\% err) from
branching ratios (BR), lifetimes ($\tau$), combined effect of 
\dEM{K\ell} and \dSU{K\ell} ($\Delta$), and phase space integrals (Int).}
\label{tab:Vusf0}
\end{table*}
\begin{sloppypar}
To carry out this analysis, values are needed for the hadronic constant
\fp , as well as \FKpi\ (see in the following). 
These values are obtained using lattice QCD, and various determinations have been 
performed. The lattice QCD community, as represented by the FlaviaNet 
Lattice Averaging Group, 
is progressing towards convergence on a set of reference values, 
but in the meantime, for the 
purposes of this work, we are led to propose our own: \FKpi=1.193(6) and \fp = 0.959(5).
Our adoption of 
these values is intended to illustrate what precision can be obtained 
in testing the SM, given current experimental and 
theoretical results. In particular, wherever possible we quote results
for quantities such as \Vusf\ or $\Vusd\times\FKpi$, which are independent
of lattice inputs and ready for use as new lattice results become available.
\end{sloppypar} 

\begin{sloppypar}
The principal experimental result of this review is the average value   
\Vusf=0.2163(5),
which has an uncertainty of about of $0.2\%$.  
The results from the five modes are in good agreement; the
fit gives $\chi^2/{\rm ndf}=0.77/4$ ($P=94\%$).
\Tab{tab:Vusf0} gives an approximate breakdown of the sources
contributing to the total uncertainty on the determination of 
\Vusf\ from each mode. 
\end{sloppypar} 
\begin{figure}
\centering
\includegraphics[width=0.4\linewidth]{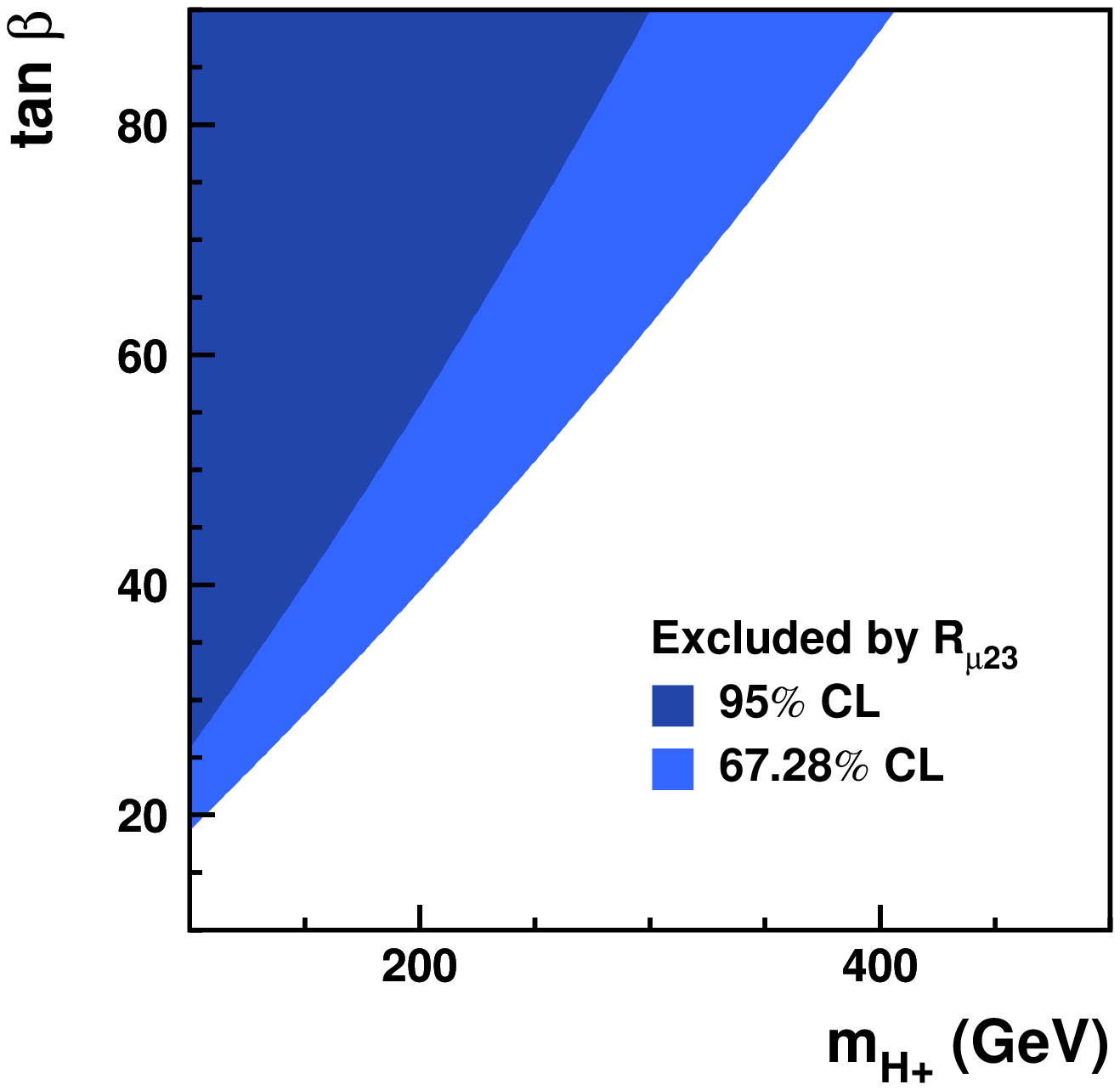}
\includegraphics[width=0.4\linewidth]{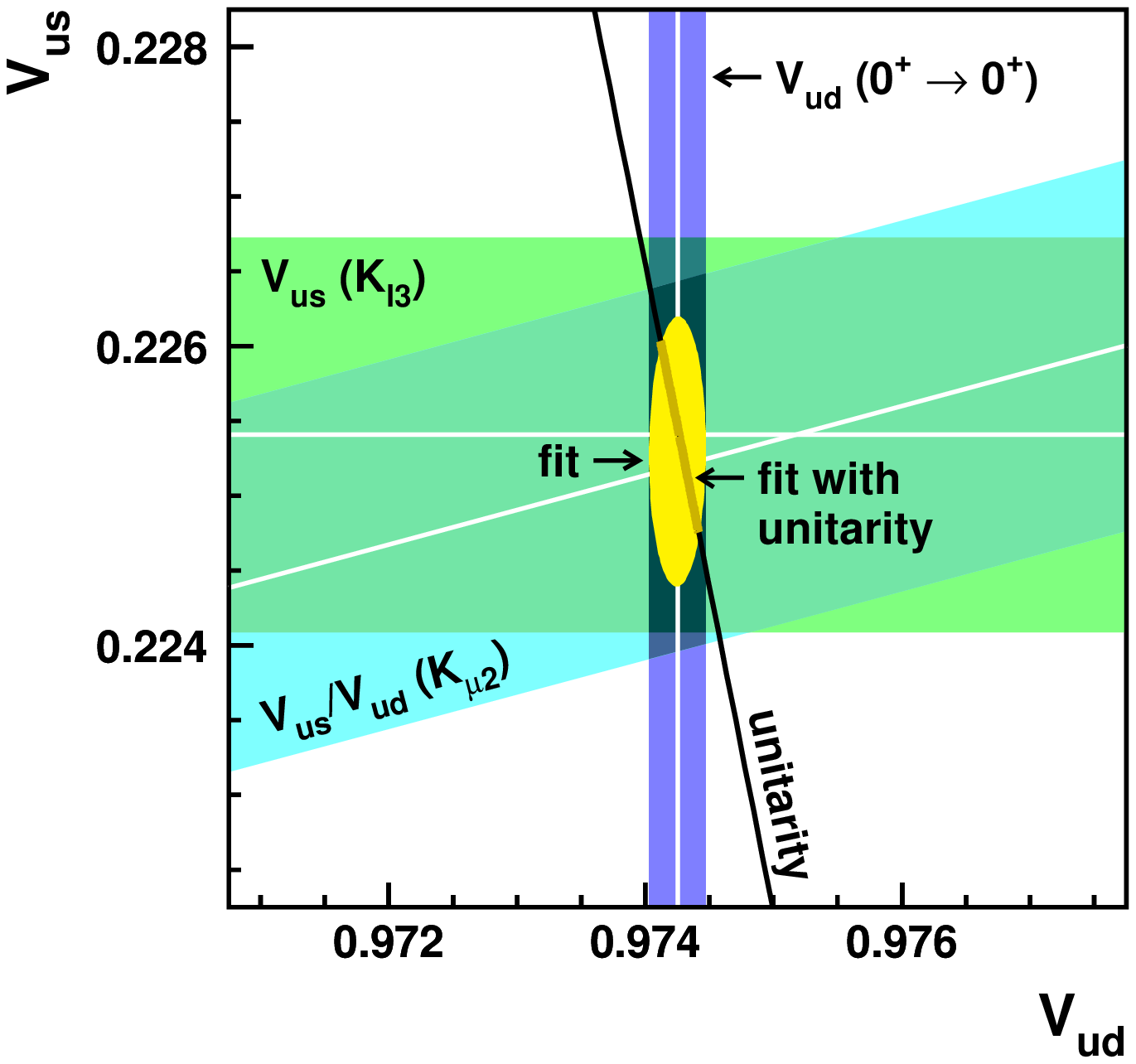}
\caption{
Left:  Results of fits to \Vud, \Vus, and \Vusd.
Right: Regions in the $(m_{H^\pm}, \tan \beta)$ plane 
in two-Higgs-doublet models excluded by the present 
result for $R_{\mu23}$. }
\label{fig:Vusf0}
\end{figure}

Equation 
$$
\frac{\Gamma_{K_{\ell2}}}{\Gamma_{\pi_{\ell2}}} =
\frac{\Vus^2}{\Vud^2}\frac{\FK^2}{\Fpi^2}
\frac{m_K(1 - m_\ell^2/m_K^2)^2}{m_\pi(1 - m_\ell^2/m_\pi^2)^2} 
\left(1 + \dEM{}\right)
$$,
allows the ratio $\Vusd\times\FKpi$ to be determined from experimental
information on the radiation-inclusive $K_{\ell2}$ and $\pi_{\ell2}$ decay 
rates. The limiting uncertainty is that from \BR{K_{\mu2(\gamma)}}, which is 
0.28\%. 
Using this, together with the value of $\tau_{K^\pm}=12.384(15)$ ns 
and 
$\Gamma(\pi^\pm\to\mu^\pm\nu)=38.408(7)~\mu{\rm s}^{-1}$, 
we obtain
$
\Vusd \times \FKpi = 0.2758(5).
$
\begin{sloppypar}
We determine \Vus\ and \Vud\ from a fit to the results obtained above.
As starting points, we use the value $\Vusf = 0.2163(5)$, 
together with the lattice QCD estimate 
$\fp = 0.959(5)$. 
We also use the result $\Vusd \times \FKpi = 0.2758(5)$ 
together with the lattice estimate
$\FKpi = 1.193(6)$. 
Thus we have
$
\Vus  = 0.2254(13) (K_{\ell 3}~{\rm only}) 
$
and
$
\Vusd = 0.2312(13) (K_{\ell 2}~{\rm only}).
$
Finally, we use the evaluation 
$\Vud = 0.97425(22)$ from a recent survey \cite{Hardy:2008gy} 
of half-life, decay-energy, and BR measurements related to 20 superallowed 
$0^+ \to 0^+$ nuclear beta decays.
Our fit to these inputs gives 
$
\Vud = 0.97425(22), \Vus = 0.2253(9) (K_{\ell3}, K_{\ell2}, 0^+\to0^+),
$
with $\chi^2/{\rm ndf} = 0.014/1$ ($P=91\%$) and negligible correlation between
\Vud\ and \Vus. 
With the current world-average value, $\Vub = 0.00393(36)$, the first-row unitarity sum is then 
$ \Delta_{\rm CKM} = \Vud^2 + \Vus^2 + \Vub^2 - 1 = -0.0001(6)$;
the result is in striking agreement with the unitarity hypothesis.
(Note that the contribution to the sum from \Vub\ is 
essentially negligible.)
As an alternate expression of this agreement, we may state a value for
$G_{\rm CKM} = G_\mu \sqrt{\Vud^2+\Vus^2+\Vub^2}$. We obtain
$G_{\rm CKM} = 1.16633(35) \times 10^{-5}~{\rm GeV}^{-2}$,
with $G_\mu = 1.166371(6) \times 10^{-5}~{\rm GeV}^{-2}$. 
\end{sloppypar}
It is also possible to perform the fit with the unitarity constraint
included, increasing by one the number of degrees of freedom. The constrained
fit gives 
$
\Vus=\sin\,\theta_C=\lambda=0.2254(6) ({\rm with~unitarity})
$
and $\chi^2/{\rm ndf}=0.024/2$ ($P=99\%$). 
This result and that obtained above without assuming unitarity are both 
illustrated in \Fig{fig:Vusf0}.
At this point, using 
$
\Delta_{\rm{CKM}} = 4 \, \left(\hat{\alpha}_{ll}^{(3)} - \hat{\alpha}_{lq}^{(3)} 
- \hat{\alpha}_{\varphi l}^{(3)} + \hat{\alpha}_{\varphi q}^{(3)} \right)
$ \cite{Cirigliano:2009wk}
and the phenomenological value 
$ \Delta_{\rm CKM}  = -0.0001(6)$,  it is possible to set bounds on the 
effective scale of the four operators that parameterize new physics
contributions to $\Delta_{\rm CKM}$. We obtain
$\Lambda > 11~{\rm TeV} (90\%~{\rm C.L.})$. 
As noted in \cite{Cirigliano:2009wk}, 
for the operators $O_{ll}^{(3)}, O_{\varphi l}^{(3)}$, and $O_{\varphi q}^{(3)}$, 
this constraint is at the same level as the
constraints from $Z$-pole measurements. 
For the four-fermion operator $O_{lq}^{(3)}$,
$\Delta_{\rm CKM}$ improves upon existing bounds from LEP2 by an order 
of magnitude.

An empirical value for the ratio 
$$
R_{\mu23} = 
\left(\frac{\FKpi}{\fp}\right)^{\!-1}\!
\left(\left|\frac{V_{us}}{V_{ud}}\right|\frac{\FK}{\Fpi}\!\right)_{\!\mu2}
\frac{\Vud_{0^+\to0^+}}{[\Vusf]_{\ell3}},
$$
can be used to exclude regions of the $(m_{H^\pm}, \tan \beta)$
parameter space in models with two Higgs doublets, such as the 
minimal supersymmetry extension of the SM \cite{Isidori:2001fv}.
Operatively, we evaluate $R_{\mu23}$ via a fit akin to that 
used to evaluate \Vusf, 
but with separate parameters accounting for 
the values of \Vus\ from $K_{\ell3}$ and $K_{\mu2}$ decays. 
The fit then has three free parameters:
the value of \Vus\ from $K_{\ell3}$ decays, the value of \Vusd\ from 
$K_{\mu2}$ decays, and the value of \Vud\ from $0^+\to0^+$ nuclear beta decays.
The input values used for \Vus\ and \Vusd\ 
include the relevant lattice constants. The contribution to
non-helicity-suppressed $K_{\ell3}$ decays from charged Higgs exchange 
is negligible, so we include as a constraint
in the fit the first-row unitarity condition on the value of \Vus\ from
$K_{\ell3}$ decays: $\Vud^2 + \Vus^2_{K_{\ell3}} + \Vub^2= 1$. 
Expressing the results of the fit in terms of \Vus\ from $K_{\ell3}$ decays and
the ratio $R_{\mu23}$, we obtain
$
\Vus = 0.2254(8) (K_{\ell3}, 0^+\to0^+, {\rm unitarity}),
R_{\mu23} = 0.999(7) (K_{\mu2}).
$
The fit gives $\chi^2/{\rm ndf}=0.0003/1$ ($P=99\%$), 
with $\rho=-0.55$ between the parameter uncertainties in the stated basis. 
The regions of the $(m_{H^\pm}, \tan \beta)$ parameter space 
in models with two Higgs doublets excluded at the $1\sigma$ and 95\% CLs 
by this result for $R_{\mu23}$ are shown as the shaded area in \Fig{fig:Vusf0}. 
Note that this result excludes the region 
at low $m_{H^\pm}$  and large  $\tan \beta$ favoured by 
$B\to\tau\nu$ \cite{Bona:2009cj}. 

As a general conclusion, we emphasize that the \order{10~{\rm TeV}}
bound on the scale of new physics, which
follows from the verification of the first-row CKM unitarity condition, 
represents one of the most stringent constraints on
physics beyond the Standard Model.


\end{document}